\documentclass[twocolumn,showpacs,preprintnumbers,amsmath,amssymb]{revtex4}

\usepackage[pdftex]{graphicx}
\usepackage{dcolumn}
\usepackage{bm}
\usepackage{color}

\newcommand{\sandw}[3]{\langle#1|#2|#3\rangle}

\newcommand{\bra}[1]{\langle #1|}
\newcommand{\ket}[1]{|#1\rangle}
\newcommand{\Tr}[0]{\textrm{Tr}}

\newcommand{\eps}[0]{\varepsilon}
\newcommand{\pphi}[0]{\varphi}
\newcommand{\rr}[0]{\vec r}
\newcommand{\rrp}[0]{\vec{r'}}

\newcommand{\de}[0]{\partial}
\newcommand{\KS}[0]{{K\!S}}
\newcommand{\eH}[0]{{eH}}
\newcommand{\exc}[0]{{exc}}

\newcommand{\half}[0]{\frac{1}{2}}

\newcommand{\footnoteremember}[2]{
\footnote{#2}
\newcounter{#1}
\setcounter{#1}{\value{footnote}}
}
\newcommand{\footnoterecall}[1]{
\footnotemark[\value{#1}]
}
\begin{document}

\preprint{APS/123-QED}

\title{Piecewise Linearity of Approximate Density Functionals Revisited: \\ Implications for Frontier Orbital Energies}

\author{Eli Kraisler}
\affiliation{Department of Materials and Interfaces, Weizmann Institute of Science, Rehovoth 76100, Israel}

\author{Leeor Kronik}
\affiliation{Department of Materials and Interfaces, Weizmann Institute of Science, Rehovoth 76100, Israel}

\date{\today}

\begin{abstract}

In the exact Kohn-Sham density-functional theory (DFT), the total energy versus the number of electrons is a series of linear segments between integer points. However, commonly used approximate density functionals produce total energies that do not exhibit this piecewise-linear behavior. As a result, the ionization potential theorem, equating the highest occupied eigenvalue with the ionization potential, is grossly disobeyed. Here, we show that, contrary to conventional wisdom, most of the required piecewise-linearity of an arbitrary approximate density functional can be restored by careful consideration of the ensemble generalization of DFT. Furthermore, the resulting formulation introduces the desired derivative discontinuity to any approximate exchange-correlation functional, even one that is explicitly density-dependent. This opens the door to calculations of the ionization potential and electron affinity even without explicit electron removal or addition. All these advances are achieved while neither introducing empiricism nor changing the underlying functional form. The power of the approach is demonstrated on benchmark systems using the local density approximation as an illustrative example.
\end{abstract}

\maketitle

Density functional theory (DFT) is a widely popular approach to the many-electron problem~\cite{PY,DG,Primer,EngelDreizler2011,Burke12,Capelle_BirdsEye}. It is based on mapping the interacting electron system into a non-interacting one. DFT is exact in principle, but the exchange-correlation (xc) density functional, $E_{xc}[n(\rr)]$ remains unknown and is always approximated in practice.

Many constraints the exact $E_{xc}[n(\rr)]$ has to satisfy have been formulated. Of particular interest here is the piecewise-linearity property: Using a zero-temperature ensemble of integer electron states~\cite{Lieb,RvL_adv}, the realm of DFT has been extended to fractional electron numbers ($N = N_0 + \alpha$, where $N_0 \in \mathbb{N}$ and $\alpha \in [0,1]$). It has been shown~\cite{PPLB82} that the total ground-state energy, $E$, is given by
\begin{equation}\label{eq.E.tot}
    E(N) = (1-\alpha)E(N_0) + \alpha E(N_0+1).
\end{equation}
An important manifestation of piecewise-linearity~\cite{PPLB82,Cohen08,MoriS08,MoriS09,SteinKronikBaer_curvatures12} is the relation between the highest occupied orbital energy, $\eps_{ho}$, and the ionization potential (IP), $I \equiv E(N_0) - E(N_0+1)$. If piecewise linearity is maintained, $\eps_{ho}=-I$, a result known as the IP theorem~\cite{PPLB82,PerdewLevy97}.

Despite the importance of piecewise-linearity, it has long been known that commonly used functional classes, such as the local density approximation (LDA), the generalized gradient approximation (GGA), or the conventional hybrid functional approximation, grossly disobey this condition. Instead, a typically convex $E(N)$ curve is obtained (see, e.g., \cite{MoriS06,SteinKronikBaer_curvatures12,Ruzsinszky07,Vydrov07,Cohen08,HaunScu10,Cohen12}) and, correspondingly, the discrepancy between $\eps_{ho}$ and $-I$ can easily be as large as a factor of two~\cite{Chan99,AllenTozer02,KueKronik08,Teale08}.

Two main approaches have emerged in response to this problem. In one approach, various correction terms are imposed on existing underlying xc-functionals~\cite{Coco05,KulikCocoMarzari06,LanyZunger09,Dabo10,Andrade11,Zheng11,Gaiduk12,Gidopoulos12}. In another, piecewise linearity is explicitly enforced in the construction of novel range-separated hybrid functionals~\cite{BaerLivSalz10,SalzBaer09,MooreSrebAutsch12,SrebAutsch12,KronikBaer_etal_JCTCrev12,Refaely12}.

The above considerations on piecewise-linearity, or lack thereof, are all based on a description of fractional-electron systems by insertion of
a density $n(\rr)$, which integrates to a fractional $N$, into a density functional developed originally for pure states. One may question whether this straightforward application is at all optimal. Indeed, Gidopoulos~\emph{et al.}~\cite{Gidopoulos02} have observed, in the context of an excited-state ensemble, that straightforward application of the Hartree term leads to an unphysical "ghost contribution". More recently, Gould and Dobson~\cite{GouldDobson12} have made similar observations of "ghost interactions" in the context of the exact-exchange (EXX) functional with fractional spin densities, and used ensemble definitions to propose an improved, linearized EXX functional.

Here, we offer an ensemble generalization of all energy terms of an arbitrary density functional, to systems with fractional $N$.
Using the simplest functional of all, the LDA, on example systems, we find that this generalization greatly reduces the problem of the energy curve convexity, significantly restores the IP theorem, and concomitantly introduces an appropriate derivative discontinuity into the xc-potential in a natural manner. All this is achieved while neither introducing empiricism nor changing the underlying functional form.

Our considerations start with the ground state of a zero-temperature interacting-electron system with fractional $N$, described by an ensemble
state $\hat \Lambda = (1-\alpha)\ket{\Psi_{N_0}}\bra{\Psi_{N_0}} + \alpha\ket{\Psi_{N_0+1}}\bra{\Psi_{N_0+1}}$, where $\ket{\Psi_{N_0+p}}$ is a (pure) many-electron ground state with $N_0+p$ electrons and $p$ is 0 or 1\footnoteremember{foot.1}{These ground states are assumed to be non-degenerate}.
The electron density is then obtained using the density operator, $\hat{n}(\rr) = \sum_i \delta(\rr - \rr_i)$, as
\begin{equation}\label{eq.n.linear}
    n(\rr) = \Tr \{ \hat \Lambda \hat{n}\} = (1-\alpha) n_0(\rr) + \alpha n_1(\rr).
\end{equation}
$n_0(\rr)$ and $n_1(\rr)$ are the densities of the interacting systems with $N_0$ and $N_0+1$ electrons, respectively. As a result, the total energy $E$ is obtained as in Eq.~(\ref{eq.E.tot}).

In the Kohn-Sham (KS) formulation of DFT, the interacting-electron system is mapped into \emph{one} KS system of non-interacting electrons
with a fractional number of particles, $N$. Therefore, its ground state must also be an ensemble state, given by $\hat \Lambda_{\KS} = (1-\alpha)\ket{\Phi_{N_0}^{(\alpha)}}\bra{\Phi_{N_0}^{(\alpha)}} + \alpha\ket{\Phi_{N_0+1}^{(\alpha)}}\bra{\Phi_{N_0+1}^{(\alpha)}}$, where $\ket{\Phi_{N_0+p}^{(\alpha)}}$ are pure KS ground states, with $N_0+p$ electrons, respectively~\cite{DG}\footnoterecall{foot.1}\footnoteremember{foot.2}{For simplicity, we use a spin-unpolarized formalism throughout. All calculations presented below were spin-polarized, but such that only one spin channel is fractionally occupied, so the ensemble is comprised of two pure states at most, as in the formalism presented}. Each 
pure ground state is described as a Slater determinant of single-electron orbitals $\{ \pphi_i^{(\alpha)} \}$, 
corresponding to the \emph{same}, $\alpha$-dependent KS potential. In contrast to the quantities $\ket{\Psi_{N_0+p}}$ and $n_p$, all quantities of the KS ensemble are $\alpha$-dependent, a fact we emphasize via the superscript $^{(\alpha)}$. Hence, in addition to the \emph{explicit} dependence of $\hat \Lambda_{\KS}$ on $\alpha$, there also exists an \emph{implicit} dependence through $\{ \pphi_i^{(\alpha)} \}$.

Similarly to Eq.~(\ref{eq.n.linear}), the KS density is obtained as $n_{\KS}^{(\alpha)}(\rr)= \Tr \{ \hat{\Lambda}_{\KS} \hat{n}\} = (1-\alpha) \rho_0^{(\alpha)}(\rr) + \alpha \rho_1^{(\alpha)}(\rr) = \sum_{i=1}^\infty g_i |\pphi_i^{(\alpha)}(\rr)|^2$, where $\rho_p(\rr) := \sandw{\Phi_{N_0+p}^{(\alpha)}}{\hat{n}}{\Phi_{N_0+p}^{(\alpha)}} = \sum_{i=1}^{N_0+p} |\pphi_i^{(\alpha)}(\rr)|^2$, and
\begin{equation}\label{eq.gi}
    g_i = \left\{
              \begin{array}{ccc}
                1      & : & i \leqslant N_0 \\
                \alpha & : & i = N_0+1       \\
                0      & : & i > N_0+1
              \end{array}
           \right.
\end{equation}
are the occupation numbers of the KS levels. While $n_{\KS}^{(\alpha)}(\rr)$ is required to equal $n(\rr)$ by construction, we stress that $\rho_p^{(\alpha)}(\rr)$ need not equal $n_p(\rr)$. Moreover, because $n_0(\rr)$, $n_1(\rr)$ and $n(\rr)$ can all be obtained independently from each other by considering systems with different $N$, Eq.~(\ref{eq.n.linear}) can be viewed as a linearity criterion for the density, complementing Eq.~(\ref{eq.E.tot}).

We now examine the ensemble properties of the Coulomb energy of the KS system, associated with the operator $\hat{W} = \half \sum_i \sum_{j \neq i} |\rr_i - \rr_j|^{-1}$. By definition~\cite{Primer}, the Coulomb functional $W = \Tr\{\hat{\Lambda}_{\KS} \hat{W} \} = W_H + W_x$ is comprised of a Hartree ($H$) and an exchange ($x$) term. Performing the $\Tr$ operation, we can express the ensemble terms $W_H$ and $W_x$ by means of the standard, pure-state definitions of the Hartree and EXX functionals (see Supplementary Material).
We obtain:
\begin{equation}\label{eq.W.H}
W_H = (1-\alpha)E_H[\rho_0^{(\alpha)}] + \alpha E_H[\rho_1^{(\alpha)}],
\end{equation}
\begin{equation}\label{eq.W.x}
W_x = (1-\alpha)E_x[\rho_0^{(\alpha)}] + \alpha E_x[\rho_1^{(\alpha)}],
\end{equation}
where as usual
\begin{equation}\label{eq.E_H.pure}
E_H[n] = \half \int \!\!\! \int d^3r d^3r' \frac{n(\rr) n(\rrp)}{|\rr - \rrp|}
\end{equation}
and
\begin{equation}\label{eq.EXX.pure}
E_x[n] = -\half \sum_{i,j=1}^\infty g_i g_j \int \!\!\! \int d^3r d^3r'  \frac{\pphi_i^*(\rrp)\pphi_j^*(\rr)\pphi_i(\rr)\pphi_j(\rrp)}{|\rr - \rrp|}.
\end{equation}
Because $E_H[n]$ is not linear in $n$, it immediately follows that the required $W_H$ of Eq.~(\ref{eq.W.H}) is \emph{not} obtained by inserting the fractional-electron density $n_\KS^{(\alpha)}$ into Eq.~(\ref{eq.E_H.pure}). A similar statement is true for $E_x[n]$ and $W_x$~\cite{GouldDobson12}. Therefore, \emph{the Hartree and EXX functionals do not retain their usual form for ensemble states}.
Instead, $W_H = E_H[n] + \Delta E_\eH[\pphi_{N_0+1}^{(\alpha)};\alpha]$ and $W_x = E_x[n] - \Delta E_\eH[\pphi_{N_0+1}^{(\alpha)};\alpha]$, where
\begin{equation}\label{eq.E_eH}
    \Delta E_\eH = \half \alpha(1-\alpha) \int \!\!\! \int d^3r d^3r' \frac{|\pphi_{N_0+1}^{(\alpha)}(\rr)|^2 |\pphi_{N_0+1}^{(\alpha)}(\rrp)|^2}{|\rr - \rrp|}
\end{equation}
is the ensemble (\emph{e}) correction.

Note that for $\alpha = 0$ or $1$, $W_H$ and $W_x$ reduce to their usual forms~(\ref{eq.E_H.pure}) and~(\ref{eq.EXX.pure}). Thus, introduction of the term $\Delta E_\eH$ does not affect the total energies of systems with an integer $N$.
In addition, even at fractional $N$ the total energy obtained for EXX calculations with no correlation should not be affected either, as $\Delta E_\eH$ appears with opposite signs in $W_H$ and $W_x$~\footnote{This complete cancelation applies to the case where only one spin-channel is fractionally occupied - see footnote~[49]. The more general case, which allows for electron migration between spin channels, has been discussed by Mori-S\'anchez \emph{et al.}~\cite{MoriS09}, and more recently by Gould and Dobson~\cite{GouldDobson12}.}.  However, the Hartree expression is usually complemented by an approximate xc-functional, $E_{xc}[n]$, that is not the EXX. Error cancelation is then not expected and, as shown below, not obtained. Trivially, an arbitrary $E_{xc}[n]$ is not linear in $n$, but it can still be made explicitly linear in $\alpha$, in the same spirit as Eqs.~(\ref{eq.W.H}),~(\ref{eq.W.x}) above, yielding:
\begin{equation}\label{eq.E_exc}
    E_\exc[n] = (1-\alpha)E_{xc}[\rho_0^{(\alpha)}] + \alpha E_{xc}[\rho_1^{(\alpha)}].
\end{equation}
(see Supplementary Material).
Note that while the dependence of $E_\exc$ on $\alpha$ is now explicitly linear, there remains an implicit non-linear dependence via the functions $\rho_p^{(\alpha)}(\rr)$. 
For the special case of the LSDA, we refer to its ensemble generalized form, using Eq.~(\ref{eq.E_exc}), as eLSDA.

Importantly, the ensemble expressions $W_H$ (Eq.~(\ref{eq.W.H})) and $E_\exc$ (Eq.~(\ref{eq.E_exc})) no longer depend explicitly on the density $n$, even for underlying functionals that are explicitly density-dependent for pure states, such as the LSDA. Ultimately, they depend on the KS orbitals (themselves a functional of $n$) via $\rho_p^{(\alpha)}(\rr)$, as well as on $\alpha$ itself. This affects the KS potential, $v_\KS$. To remain within the KS framework, it must now be evaluated using the optimized effective potential (OEP) procedure, appropriate for implicitly density-dependent functionals~\cite{Grabo_MolPhys,KueKronik08,KumPer03a,KumPer03b,KLI}. A complete derivation of $v_\KS$ is provided in the Supplementary Material. One unusual aspect of it, which we stress here, is that the explicit dependence of $W_H$ and $E_\exc$ on $\alpha$ contributes a spatially-uniform but $\alpha$-dependent term to $v_\KS$, given by
\begin{align}\label{eq.v0}
              &v^{(0)} = -\half \int \!\! \int \frac{|\pphi_{N_0+1}^{(\alpha)}(\rr)|^2 |\pphi_{N_0+1}^{(\alpha)}(\rrp)|^2}{|\rr - \rrp|}d^3r d^3r' \\
\nonumber     &+ E_{xc}[\rho_1^{(\alpha)}] - E_{xc}[\rho_0^{(\alpha)}] - \int |\pphi_{N_0+1}^{(\alpha)}(\rr)|^2 v_{xc}[\rho_1^{(\alpha)}] d^3r,
\end{align}
where $v_{xc}=\delta E_{xc}/\delta n$ is the usual xc-potential. This term involves the highest (possibly partially) occupied orbital, $\pphi_{N_0+1}^{(\alpha)}$, and \emph{does not vanish even when $N$ is an integer}, despite the fact that for integer values the conventional and ensemble-generalized energy expressions are identical. Such a constant term, although allowed by the Hohenberg-Kohn theorem~\cite{HK'64}, is usually deemed unimportant because it does not affect the density or the total energy. However, it does shift the KS eigenvalues, a fact we show below to be crucial. Thus, all calculations now conceptually involve orbital-dependent functionals, although for integer $N$ the term $v^{(0)}$ can be easily evaluated without performing the computationally demanding OEP calculation.

\begin{figure}
  \includegraphics[scale=0.35]{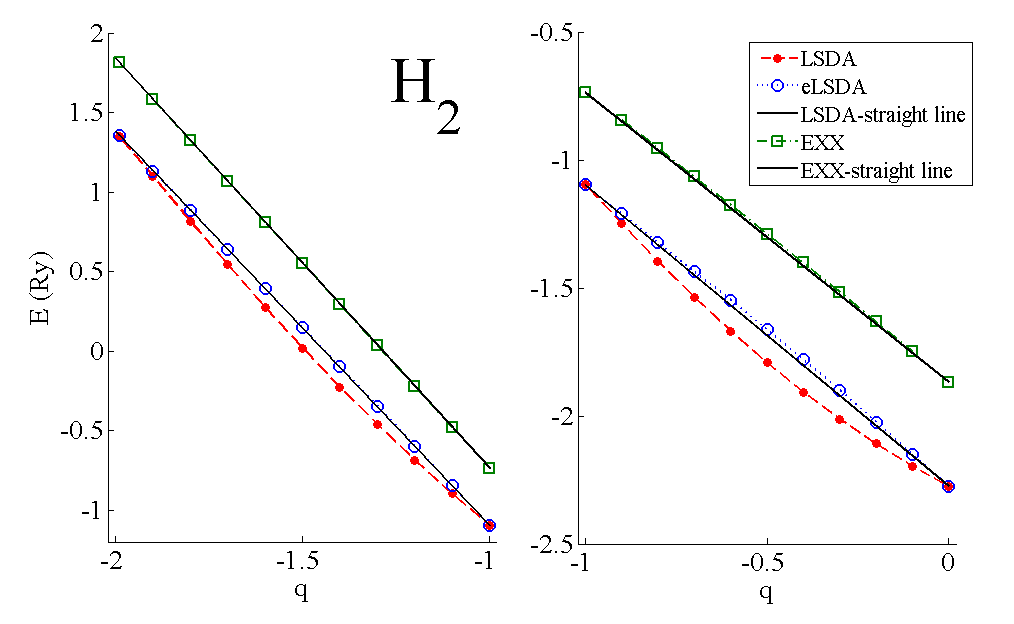}\\
  \includegraphics[trim=0mm 0mm 15mm -10mm,scale=0.35]{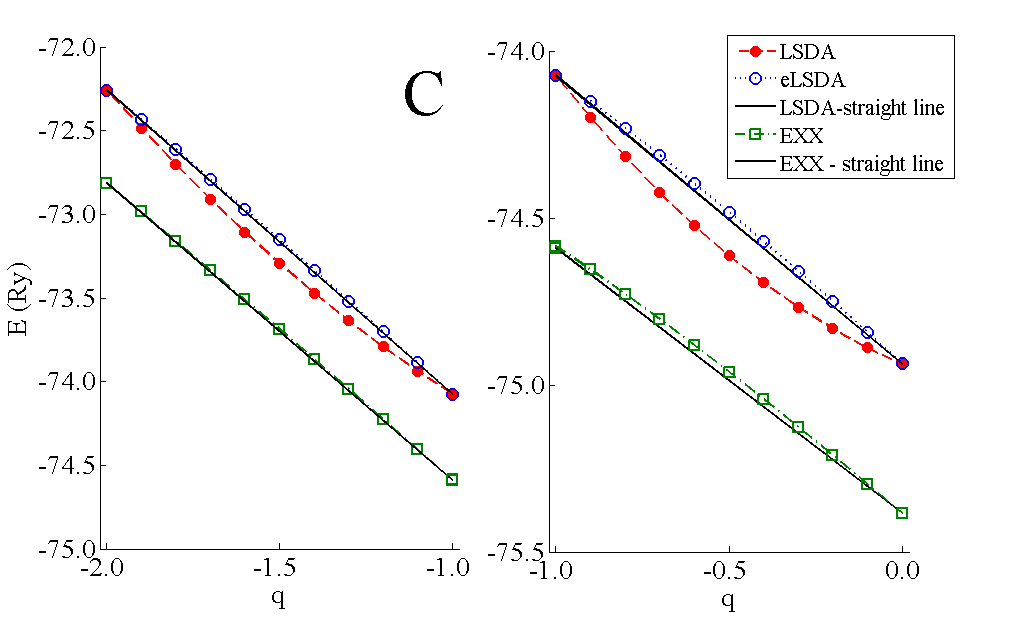} \\
  \caption{Energy of the H$_2$ molecule (top) and of the C atom (bottom) as a function of fractional charge $q$, for various functionals. EXX results for H$_2$ have been shifted upwards by 0.4 Ry, for clarity. The straight solid lines connect the energies obtained at the integer value, as a reference for complete piecewise-linearity.} 
  \label{fig.E_tot}
\end{figure}

To illustrate the proposed generalization and its implications, we apply the eLSDA functional to the H$_2$ molecule and the C atom using DARSEC -- an all-electron, real-space code~\cite{Makmal09JCTC} (numerical details are given in the Supplementary Material). The total energies for the above two systems, as a function of the net charge, $q$, are given in Fig.~\ref{fig.E_tot}, with $q$ ranging from -2 (doubly-ionized system) to 0 (neutral system). The LSDA energy curves are, as expected, convex~\cite{MoriS06,Vydrov07,Cohen08,Cohen12}. The curve for the eLSDA is, however, almost piecewise linear, being slightly concave.
The strong reduction in the deviation from piecewise-linearity is a significant advantage of the ensemble approach. This deviation is not fully eliminated because, while the eLSDA functional is explicitly linear in $\alpha$ by construction, it may still be implicitly non-linear through $\{ \pphi_i^{(\alpha)} \}$.
Comparison of the eLSDA results to the EXX ones shows that the piecewise-linearity of eLSDA is comparable to that of EXX. An obvious advantage of eLSDA, however, is the treatment of correlation.

\begin{figure}
  \includegraphics[trim= 0mm 0mm 20mm 0mm,scale=0.32]{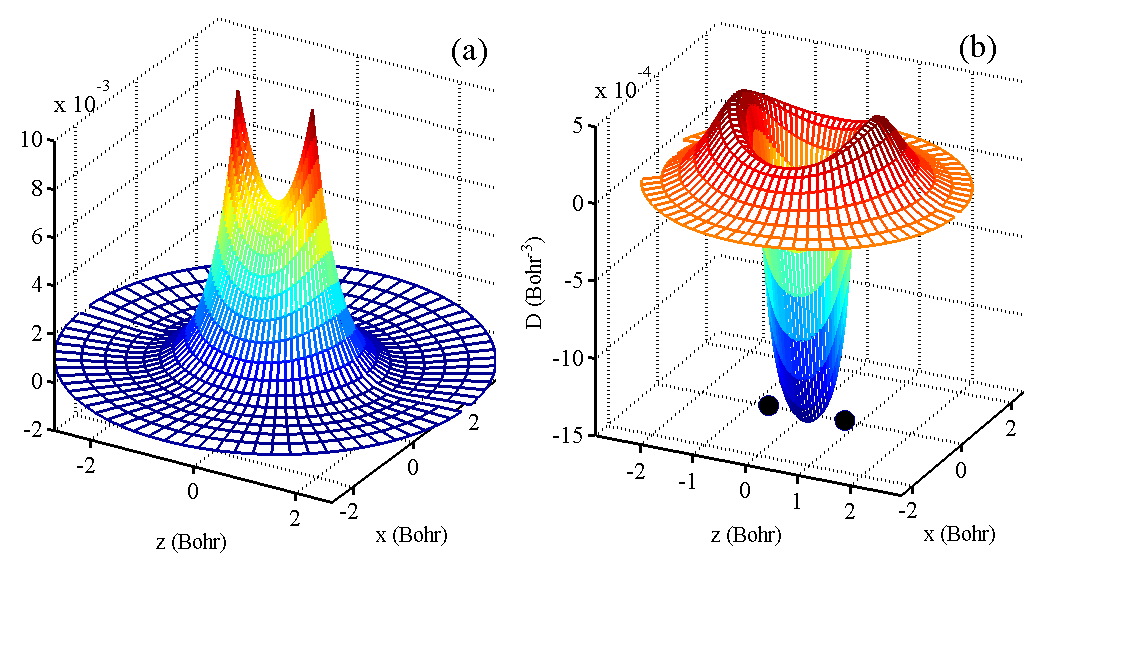}
  \caption{Deviation from piecewise-linearity in the density, $D(\rr)$, obtained for the H$_2$ molecule for $q=-0.5$ using (a) LSDA, (b) eLSDA}\label{fig.D.LSDA}
\end{figure}

eLSDA affords a significant improvement also in satisfying the density linearity criterion, Eq.~(\ref{eq.n.linear}).
We consider $D(\rr) := n(\rr) - (1-\alpha) n_0(\rr) - \alpha n_1(\rr)$, which should equal 0 at all $\rr$ for the exact functional. A plot of $D(\rr)$ at $q=-0.5$ for H$_2$, as obtained with LSDA and eLSDA, is presented in Fig.~\ref{fig.D.LSDA}. Clearly, the spatial profile of $D(\rr)$ is smoother with eLSDA and its average numerical value much smaller. Specifically, $Q(q):=\int D^2(\rr) d^3r$, which is the variance of $D(\rr)$ per a given $q$, is $\sim 10^{-4}$ Bohr$^{-3}$ with LSDA. With eLSDA, however, it is lower by two orders of magnitude for $q=-1...0$ and essentially zero for $q=-2...-1$.

\begin{figure}
  \includegraphics[width=8.0cm]{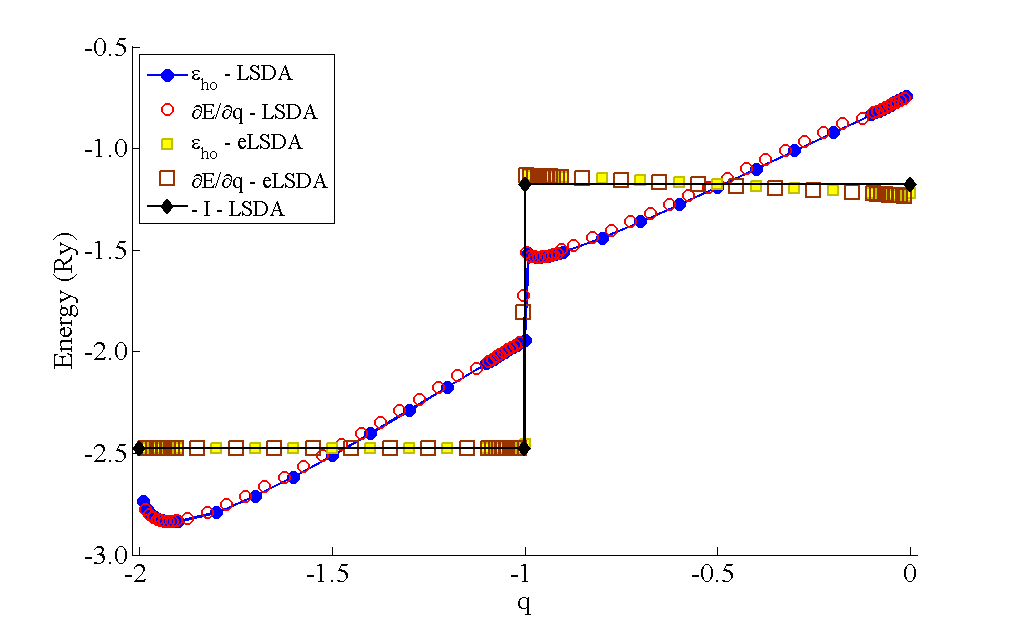}\\
  \caption{Frontier orbital energy, $\eps_{ho}$, energy derivative $\de E/\de q$, as a function of $q$, and the negative of the IP, $-I$, calculated for H$_2$ with the LSDA and eLSDA functionals}\label{fig.eps.f}
\end{figure}

The great improvement in the piecewise-linearity of the energy curve (Fig.~(\ref{fig.E_tot})) is directly manifested in the degree to which the IP theorem is satisfied. This is illustrated in Fig.~\ref{fig.eps.f}. The figure shows the highest (possibly partially) occupied orbital,
$\eps_{ho}$, the energy derivative $\de E/\de q$, and the negative of the IP, $-I$ (computed from total energy differences obtained at integer $q$ values), as calculated for H$_2$ as a function of $q$ with both LSDA and eLSDA. Janak's theorem~\cite{Janak}, which equates between $\eps_{ho}$ and $\de E/\de q$ for any approximate functional, is indeed closely obeyed by both approximations.
But because eLSDA is much more piecewise-linear, $\eps_{ho}(q)$ calculated with it is much more piecewise-constant as a function of $q$ (as it should be for the exact functional). Furthermore, $\eps_{ho}$ coincides much more closely with $-I$ when approaching an integer $q$ from below, in agreement with the IP theorem~\cite{PPLB82,PerdewLevy97} -- a direct consequence of the constant potential $v^{(0)}$.

\begin{table}
  \centering
  \begin{tabular}{llrrrr}
    \hline
            &               & LSDA  & eLSDA & EXX   & experiment\footnote{Ref.~\cite{HandChemPhys92}} \\
    \hline
  H$_2$     & $-\eps_{ho}$  & 0.745 & 1.223 & 1.193 &            \\
            & $I$           & 1.178 & 1.178 & 1.130 & 1.134      \\
            & $\Delta_{ho}$ & 37\%  & 4\%   &  6\%  &            \\
  H$_2^+$   & Gap           & 0.426 & 1.320 & 1.489 &            \\
            & $I_2 - I_1$   & 1.298 & 1.298 & 1.446 & 1.443\footnote{For H$_2^+$, no experimental value for $I_2$ exists. Instead, it was obtained from EXX calculations, which yield an exact result for this system.}      \\
            & $\Delta_{gap}$& 71 \% &  2\%  & 3 \%  &            \\

    \hline
  C         & $-\eps_{ho}$  & 0.450 & 0.942 & 0.876 &            \\
            & $I$           & 0.859 & 0.859 & 0.793 & 0.828      \\
            & $\Delta_{ho}$ & 48\%  & 10\%  & 10\%  &            \\
  C$^+$     & Gap           & 0.019 & 1.125 & 1.140 &            \\
            & $I_2 - I_1$   & 0.962 & 0.962 & 0.982 & 0.965      \\
            & $\Delta_{gap}$& 98 \% & 17 \% & 16 \% &            \\
    \hline

    \end{tabular}
  \caption{Highest occupied orbital energy, $-\eps_{ho}$, compared to the IP, $I$; Cation fundamental gap, deduced from the discontinuity of $\eps_{ho}$ at $q=-1$, compared to the difference between the second and first IP's of the neutral system. All quantities are computed for both H$_2$ and C and all energies are given in Ry. $\Delta$'s correspond to the relative error between the two values positioned immediately above them.}\label{table.eps.f}
\end{table}

The satisfaction of the IP theorem is closely related to another fundamental property of the exact xc-functional: as the number of electrons crosses an integer, the xc-potential may ``jump'' by a constant, usually known as the derivative discontinuity (DD)~\cite{PPLB82}. The conventional wisdom on explicit density functionals (including LSDA) is that they do not possess this discontinuity.
Recently, Stein \emph{et al.}~\cite{SteinKronikBaer_curvatures12} have shown that a significant increase in the degree of piecewise-linearity \emph{must} be accompanied by the appearance of a discontinuity in the xc-potential. Here it emerges from $v^{(0)}$ of Eq.~(\ref{eq.v0}),
which depends on the highest occupied orbital and is therefore different if one approaches an integer $N$ from the left or from the right.
Therefore, the DD of explicit density functionals arises naturally, without invoking any empiricism.
This is readily observed in Fig.~\ref{fig.eps.f} and Table~\ref{table.eps.f}: the fundamental gap of the ion H$_2^+$, deduced from the discontinuity in $\eps_{ho}$ around $q=-1$, is much larger with eLSDA than with LSDA, and corresponds much more closely to the result obtained from total energy differences (solid black line in the figure). Similar observations apply to the C atom (see Table).
Thus, our ensemble-based approach automatically identifies and restores the missing derivative discontinuity, appropriate for any underlying functional. Importantly, as the potential ``jumps'' by a constant at the integer-electron point, owing to the derivative discontinuity, the KS orbitals do not change at all. Therefore, the missing derivative discontinuity can be evaluated using only the Kohn-Sham eigenvalues and orbitals of the integer point itself. This opens the door to calculations of the ionization potential and electron affinity even without explicit electron removal or addition.

In conclusion, we presented a generalization of the Hartree, exchange and correlation terms of an arbitrary density functional to systems with a fractional electron number, based on the ensemble form of DFT. Using the local density approximation on H$_2$ and C as illustrative examples, we showed that this generalization significantly reduces the deviation from piecewise linearity and generates the appropriate derivative discontinuity, without introducing empiricism and with no changes to the underlying functional form. With this generalization, the total energy at integer electron numbers remains intact, but the eigenenergies change and the IP theorem is much more closely obeyed. This shows that problems that have plagued simple approximate density functionals for many years can be very strongly mitigated by rigorous employment of ensemble DFT within the OEP approach, without any further functional development. We expect this proposed generalization to be equally useful for more advanced approximate functionals, as well as for more complex systems, allowing for improvement in spectroscopic properties without any compromise on energetics.

Work was supported by the European Research Council, the Germany-Israel Science Foundation, and the Lise Meitner - Minerva Center for Computational Chemistry. E.K. acknowledges the help of Shira Weissman (Weizmann Institute) and fruitful discussions with Tobias Schmidt and Stephan K\"ummel (Bayreuth University).

\bibliography{bibliography}

\end{document}


\title{SUPPLEMENTARY MATERIAL for: \\
\emph{``The piecewise-linearity of approximate density functionals revisited: \\
implications for frontier orbital energies''}}

\author{Eli Kraisler}
\affiliation{Department of Materials and Interfaces, Weizmann Institute of Science, Rehovoth 76100, Israel}

\author{Leeor Kronik}
\affiliation{Department of Materials and Interfaces, Weizmann Institute of Science, Rehovoth 76100, Israel}

\date{\today}


\maketitle


\textbf{Contents}

\begin{enumerate}[I.]
  \item Derivation of Eqs.~(\eqmWH), (\eqmWx), (\eqmEeH)
  \item Rationale of Eq.~(\eqmEexc)
  \item Derivation of the KS potential
  \item Numerical details on calculation of the systems \\ H$_2$ and C
\end{enumerate}

In the following, reference to equations from the main text are made by a single number (e.g.\ Eq.~(1)), while equations from this document include the section number, as well (e.g.\ Eq.~(I.1)).


\section{Derivation of Eqs.~(\eqmWH), (\eqmWx), (\eqmEeH)}

To obtain Eqs.~(\eqmWH) and~(\eqmWx), consider the application of the Coulomb operator
\begin{equation}
    \hat{W} = \half \sum_{i=1}^N \sum_{\substack{j=1 \\ j \neq i}}^N \frac{1}{|\rr_i - \rr_j|}
\end{equation}
to a pure state $\ket{\Phi}$ of the KS system, where
\begin{equation} \label{Phi.Single.Slater.det}
\ket{\Phi} = \frac{1}{\sqrt{N!}} \begin{vmatrix}
    \pphi_1(\vec{r_1}) & \pphi_1(\vec{r_2}) & ... & \pphi_1(\vec{r_N}) \\
    \pphi_2(\vec{r_1}) & \pphi_2(\vec{r_2}) & ... & \pphi_2(\vec{r_N}) \\
    \vdots & \vdots & \ddots & \vdots \\
    \pphi_N(\vec{r_1}) & \pphi_N(\vec{r_2}) & ... & \pphi_N(\vec{r_N}) \\
\end{vmatrix}
\end{equation}
is a Slater determinant of the one-electron KS orbitals $\{ \pphi_i \}$. The symbol $^{(\alpha)}$ does not accompany here the KS orbitals and other derived quantities, as opposed to the main text, for clarity of presentation.

The quantity $W:=\sandw{\Phi}{\hat W}{\Phi}$ is given by the well-known expression
\begin{eqnarray}
\nonumber W&=&\half\sum_{i=1}^N \sum_{j=1}^N \int \!\!\! \int d^3r d^3r' \frac{\pphi_i^*(\rr)\pphi_i(\rr)\pphi_j^*(\rrp)\pphi_j(\rrp)}{|\rr-\rrp|} \\
\nonumber  &-&\half\sum_{i=1}^N \sum_{j=1}^N \int \!\!\! \int d^3r d^3r' \frac{\pphi_i^*(\rrp)\pphi_j^*(\rr)\pphi_i(\rr)\pphi_j(\rrp)}{|\rr - \rrp|} \\
\end{eqnarray}
In the first term above we recognize the pure-state density, defined as $\rho=\sum_{i=1}^N |\pphi_i|^2$, and partition the Coulomb energy as
\begin{equation}
    W = E_H[\rho] + E_x[\rho],
\end{equation}
where
\begin{equation}\label{eq.E_H.pure}
E_H[\rho] = \half \int \!\!\! \int d^3r d^3r' \frac{\rho(\rr) \rho(\rrp)}{|\rr - \rrp|}
\end{equation}
is the Hartree energy, and
\begin{equation}\label{eq.EXX.pure}
E_x[\rho] = -\half \sum_{i=1}^N \sum_{j=1}^N \int \!\!\! \int d^3r d^3r'  \frac{\pphi_i^*(\rrp)\pphi_j^*(\rr)\pphi_i(\rr)\pphi_j(\rrp)}{|\rr - \rrp|}
\end{equation}
is the exchange energy (where $\pphi_i$ themselves are functionals of the density).

For an ensemble state, $W$ is obtained using the $\Tr$ procedure: $W = \Tr\{\hat{\Lambda}_{\KS} \hat{W} \}$, where
\begin{equation}
    \hat \Lambda_{\KS} = (1-\alpha)\ket{\Phi_{N_0}}\bra{\Phi_{N_0}} + \alpha\ket{\Phi_{N_0+1}}\bra{\Phi_{N_0+1}},
\end{equation}
and
\begin{equation}
    W = (1-\alpha)\sandw{\Phi_{N_0}}{\hat W}{\Phi_{N_0}} + \alpha\sandw{\Phi_{N_0+1}}{\hat W}{\Phi_{N_0+1}}.
\end{equation}
The densities corresponding to the states $\ket{\Phi_{N_0}}$ and $\ket{\Phi_{N_0+1}}$ are $\rho_0$ and $\rho_1$, respectively (see definitions in the main text), and therefore
\begin{equation}
    W = (1-\alpha)E_H[\rho_0] + \alpha E_H[\rho_1] + (1-\alpha)E_x[\rho_0] + \alpha E_x[\rho_1].
\end{equation}
The first two terms are the Hartree energy of the ensemble state, denoted by $W_H$. The last two terms are the exchange energy of the ensemble state, denoted by $W_x$. These four terms appear in Eqs.~(\eqmWH) and~(\eqmWx).

To express $W_H$ in terms of $E_H[n]$ and $\Delta E_{eH}$, we insert the definition $n = (1-\alpha)\rho_0 + \alpha \rho_1$ (Eq.~(\eqmnlinear)) into Eq.~(\ref{eq.E_H.pure}) to obtain
\begin{align}
\nonumber &E_H[n] &&= (1-\alpha)^2 E_H[\rho_0] + \alpha^2 E_H[\rho_1] \\
\nonumber &       &&  \hspace{20mm}+\alpha(1-\alpha)\int \!\!\! \int d^3r d^3r' \frac{\rho_0(\rr) \rho_1(\rrp)}{|\rr - \rrp|} \\
\nonumber &       &&= (1-\alpha) E_H[\rho_0] + \alpha E_H[\rho_1] &
\end{align}
\begin{equation}\label{eq.E_H.interm}
+\alpha(1-\alpha) \lp( \int \!\!\! \int d^3r d^3r' \frac{\rho_0(\rr) \rho_1(\rrp)}{|\rr - \rrp|} - E_H[\rho_0] - E_H[\rho_1] \rp)
\end{equation}

Combining Eqs.~(\ref{eq.E_H.pure}) and~(\ref{eq.E_H.interm}) while using the fact that $\rho_1 - \rho_0 = |\pphi_{N_0+1}|^2$, yields
\begin{align}
\nonumber &E_H[n] = (1-\alpha) E_H[\rho_0] + \alpha E_H[\rho_1] \\
\nonumber &-\half \alpha(1-\alpha) \int \!\!\! \int d^3r d^3r' \frac{|\pphi_{N_0+1}(\rr)|^2 |\pphi_{N_0+1}(\rrp)|^2}{|\rr - \rrp|}.\\
\end{align}
Thus, we arrive at the relation $W_H = E_H[n] + \Delta E_\eH[\pphi_{N_0+1};\alpha]$.

To express $W_x$ in terms of $E_x[n]$ and $\Delta E_{eH}$, we rearrange Eq.~(\eqmWx) as
\begin{equation}\label{eq.W_x.n.interm1}
W_x = E_x[\rho_0] + \alpha (E_x[\rho_1]-E_x[\rho_0]).
\end{equation}
Using Eq.~(\ref{eq.EXX.pure}) one then obtains
\begin{align}\label{eq.E_x.rho1-E_x.rho0}
\nonumber &E_x[\rho_1]-E_x[\rho_0] = &\\
    &-\half \sum_{j=1}^{N_0+1} \int \!\!\! \int d^3r d^3r'  \frac{\pphi_{N_0+1}^*(\rrp)\pphi_j^*(\rr)\pphi_{N_0+1}(\rr)\pphi_j(\rrp)}{|\rr - \rrp|}
\end{align}
From Eq.~(\eqmgi) $g_{N_0+1} = \alpha$; considering Eq.~(\ref{eq.W_x.n.interm1}), while substituting Eq.~(\ref{eq.EXX.pure}) for its first term and Eq.~(\ref{eq.E_x.rho1-E_x.rho0}) for the second term, we obtain
\begin{align}
\nonumber    W_x = &-\half \sum_{i=1}^\infty \sum_{j=1}^{N_0} g_i \int \!\!\! \int d^3r d^3r'  \frac{\pphi_i^*(\rrp)\pphi_j^*(\rr)\pphi_i(\rr)\pphi_j(\rrp)}{|\rr - \rrp|} \\
    &-\half \alpha \int \!\!\! \int d^3r d^3r'  \frac{|\pphi_{N_0+1}(\rrp)|^2 |\pphi_{N_0+1}(\rr)|^2}{|\rr - \rrp|}
\end{align}
Finally, to achieve the form of Eq.~(\eqmEx), we replace the sum $\sum_{j=1}^{N_0}$ by $\sum_{j=1}^\infty g_i$. This change introduces an additional term to the double summation, which should be subtracted for maintaining the equality. This manipulation leads to the expression
\begin{align}
\nonumber    W_x = &-\half \sum_{i=1}^\infty \sum_{j=1}^\infty g_i g_j \int \!\!\! \int d^3r d^3r'  \frac{\pphi_i^*(\rrp)\pphi_j^*(\rr)\pphi_i(\rr)\pphi_j(\rrp)}{|\rr - \rrp|} \\
    &-\half \alpha(1-\alpha) \int \!\!\! \int d^3r d^3r' \frac{|\pphi_{N_0+1}(\rrp)|^2 |\pphi_{N_0+1}(\rr)|^2}{|\rr - \rrp|},
\end{align}
which is equivalent to the form $W_x = E_x[n] - \Delta E_\eH[\pphi_{N_0+1};\alpha]$.

\section{Rationale of Eq.~(\eqmEexc)}

In the following we explain in more detail the rationale behind the ensemble generalization of an approximate xc-functional to a form which is explicitly linear in $\alpha$.

The approximate xc-functional can be (and usually is) presented as a sum of an approximate exchange functional and an approximate correlation functional. Because the exact exchange functional is explicitly linear in $\alpha$, as shown in Eq.~(\eqmWx), it is reasonable to require that the approximate exchange functional exhibit the same property.

The correlation functional for a pure state can be formally expressed, without any approximation, as (see e.g.\ Eq.(1.68) in Ref.~\cite{Primer})
\begin{equation}
    E_c = \sandw{\Psi}{\hat{T}+\hat{W}}{\Psi} - \sandw{\Phi}{\hat{T}+\hat{W}}{\Phi},
\end{equation}
where $\hat{T}$ and $\hat{W}$ are the kinetic and the Coulomb operators, respectively, $\Psi$ is the wavefunction of the interacting system, and $\Phi$ is the wavefunction of the KS system.

For ensembles, the wavefunctions $\Psi$ and $\Phi$ of the pure state are substituted by the ensemble operators 
$\hat{\Lambda}$ and $\hat{\Lambda}_\KS$, respectively (see definitions in the main text). The correlation energy then becomes
\begin{align}
\nonumber    E_c = (1-\alpha) \sandw{\Psi_{N_0}}{\hat{T}+\hat{W}}{\Psi_{N_0}} + \alpha \sandw{\Psi_{N_0+1}}{\hat{T}+\hat{W}}{\Psi_{N_0+1}}\\
-(1-\alpha) \sandw{\Phi_{N_0}^{(\alpha)}}{\hat{T}+\hat{W}}{\Phi_{N_0}^{(\alpha)}} - \alpha \sandw{\Phi_{N_0+1}^{(\alpha)}}{\hat{T}+\hat{W}}{\Phi_{N_0+1}^{(\alpha)}},
\end{align}
being explicitly linear in $\alpha$. Thus, it is reasonable to require that the approximate correlation functional will also be explicitly linear in $\alpha$.

Explicit linearization in $\alpha$ of the exchange-correlation functional carries with it an additional advantage. Any successful underlying density functional obeys various exact constraints. In particular, in LSDA a sum rule for the exchange-correlation hole is obeyed \cite{DG,PY}. Because the pure-state functional is then generalized into an ensemble one via an appropriate linear combination, exact constraints obeyed by the integer-electron functional will automatically be carried over to the ensemble-generalized functional.

\section{Derivation of the KS potential}

In our ensemble approach, the KS potential, $v_\KS$, cannot be obtained by a straightforward application of a functional derivative, because parts of the energy functional $E[n] = T_\KS + V_n[n] + E_H[n] + \Delta E_\eH [\pphi_{N_0+1}^{(\alpha)};\alpha] + E_\exc[\rho_0^{(\alpha)},\rho_1^{(\alpha)};\alpha]$ depend explicitly on the orbitals and $\alpha$, rather than on $n$.

Let us denote $v_\KS = v_n + v_H + v_\eH + v_\exc$, where $v_H = \delta E_H/\delta n$, $v_\eH = \delta (\Delta E_\eH)/\delta n$ and $v_\exc = \delta E_\exc/ \delta n$. The two last terms of $v_\KS$ can be presented as
\begin{equation}\label{eq.v_T}
    v_T(\rr) = \frac{\delta E_T
    }{\delta n} = \lp( \frac{\de E_T}{\de \alpha} \rp)_n \frac{\delta \alpha}{\delta n} + \lp( \frac{\delta E_T}{\delta n(\rr)} \rp)_\alpha,
\end{equation}
where $T$ stands for either the $\eH$ term or $\exc$ term. For the ensemble of two pure, non-degenerate states discussed throughout, $\alpha[n] = N - {\rm floor}(N)$ and $N=\int n d^3r$. We then find $\delta \alpha / \delta n = 1$. The second term of the RHS of Eq.~(\ref{eq.v_T}) is denoted by $v_T^{(1)}$ and can be expressed as
\begin{equation}
    v_T^{(1)}(\rr) = \sum_{i=1}^\infty \int d^3r' \lp( \frac{\delta \pphi_i(\rrp)}{\delta n(\rr)} \rp)_\alpha \lp( \frac{\delta E_T}{\delta \pphi_i(\rrp)} \rp)_\alpha + c.c.,
\end{equation}
where $\alpha$ is constant and therefore treated as a parameter. Because $\delta E_T/\delta \pphi_i$ can be obtained from Eqs.~(\eqmEeH) and~(\eqmEexc), $v_T^{(1)}$ can be found using the OEP procedure~\cite{KueKronik08}, as is suitable for an orbital-dependent functional.

The first term of the RHS of Eq.~(\ref{eq.v_T}) is denoted by $v_T^{(0)}$. This term is somewhat unusual, as most functionals do not contain the quantity $\alpha$ explicitly. $v_T^{(0)}$ is $\alpha$-dependent, but constant in space. As the expressions for $\Delta E_\eH$ and $E_\exc$ do not depend explicitly on the density $n$, it is not possible to take the derivative $(\de E_T/\de \alpha)_n$ directly. However, the quantity $(\de E_T/\de \alpha)_{\{\pphi_i\}}$ is accessible. To relate between the two quantities, we express $E_T$ as a functional of $\alpha$ and $n$, where the latter is itself a functional of $\alpha$ and $\{ \pphi_i \}$: $E_T = E_T \lp[ \alpha, n[ \{ \pphi_i \}, \alpha ] \rp]$. Then, we obtain
\begin{equation}
    \lp( \frac{\de E_T}{\de \alpha} \rp)_{\{\pphi_i\}} = \lp( \frac{\de E_T}{\de \alpha} \rp)_n + \int d^3r \lp( \frac{\delta E}{\delta n(\rr)} \rp)_\alpha \lp(\frac{\de n(\rr)}{\de \alpha}\rp)_{\{\pphi_i\}}.
\end{equation}
On the right-hand side of this expression we recognize the first term to be $v_T^{(0)}$ and $v_T^{(1)}$ to be the first multiplicand of the second term. From Eq.~(\eqmnlinear), $(\de n/\de \alpha)_{\{\pphi_i\}} = |\pphi_{N_0+1}^{(\alpha)}|^2$. As a result,
\begin{equation}\label{eq.v_T.0}
    v_T^{(0)} = \lp( \frac{\de E_T}{\de \alpha} \rp)_{\{\pphi_i\}} - \int |\pphi_{N_0+1}^{(\alpha)}(\rr)|^2 v_T^{(1)}(\rr) d^3r.
\end{equation}
We remind that in the OEP formalism for the highest occupied orbital
\begin{align}
\nonumber    \bar{v}^{(1)}_{T,N_0+1} := &\int |\pphi_{N_0+1}^{(\alpha)}(\rr)|^2 v_T^{(1)}(\rr) d^3r =\\
    = &\int |\pphi_{N_0+1}^{(\alpha)}(\rr)|^2 u_{T,N_0+1}(\rr) d^3r =: \bar{u}_{T,N_0+1},
\end{align}
where
\begin{equation}
u_{T,N_0+1} = \frac{1}{g_{N_0+1}} \frac{1}{\pphi_{N_0+1}^{* (\alpha)}} \frac{\delta E_T}{\delta \pphi_{N_0+1}}
\end{equation}
(see~\cite{KueKronik08}, Sec.II and~\cite{Krieger92}). An analytical form for $v_T^{(0)}$ can then be obtained from Eqs.~(\eqmEeH) and~(\eqmEexc). Using this relation, for the $\eH$ functional we find:
\begin{equation}\label{eq.v0.fh}
v_\eH^{(0)} = - \half \int \!\! \int \frac{|\pphi_{N_0+1}^{(\alpha)}(\rr)|^2 |\pphi_{N_0+1}^{(\alpha)}(\rrp)|^2}{|\rr - \rrp|} d^3r d^3r',
\end{equation}
and for the $\exc$-functional:
\begin{equation}\label{eq.v0.exc}
v_\exc^{(0)} = E_{xc}[\rho_1^{(\alpha)}] - E_{xc}[\rho_0^{(\alpha)}] - \int |\pphi_{N_0+1}^{(\alpha)}(\rr)|^2 v_{xc}[\rho_1^{(\alpha)}] d^3r.
\end{equation}

During the self-consistent numerical solution of the KS equations with the proposed functionals, the spatial constant $v_T^{(0)}$ can be omitted, as the addition of a constant to the potential does not affect the eigenfunctions, the density, or the total energy $E$. However, $v_T^{(0)}$ has to be taken into account when addressing KS eigenenergies, in particular when comparing the energy of the frontier orbital to the derivative $\de E/\de q$ (see definition in the main text), following the Janak and the IP theorems~\cite{PPLB82,Janak}, and when calculating the energy gap as $v_T^{(0)}$ varies with $N_0$ and thus does not cancel out.

The fact that a constant shift in the KS potential does not lead to a shift in the total energy of the system can be viewed in two different, equivalent ways. In one way, the total energy in DFT is expressed by definition as a sum of the Kohn-Sham kinetic energy, the electron-ion energy, the Hartree energy, and the exchange-correlation energy (see, e.g. Ref.\ \cite{PY}, Eq. (7.2.1)). An additive constant in the potential does not affect the KS orbitals, which are obtained from solving the KS equation. Therefore, the density obtained from these orbitals is not affected either. Because all the energy ingredients, i.e. the kinetic, the ion-electron, Hartree, and xc-energy, depend only on the density and/or the KS orbitals, the energy does not change. Alternatively, the total energy in DFT can be obtained using the KS eigenenergies, as in Eq. (7.2.10) in Ref.\ \cite{PY}. In this approach, all eigenenergies are shifted by the same arbitrary constant $C$, which changes the total energy by $C \cdot N$, where $N$ is the number of electrons. However, this is then compensated by the change in the last term of
Eq.\ (7.2.10), which includes the difference between the xc-energy and the integral over the density-weighted xc-potential. Because the xc-potential shifts by the same constant $C$, this term yields a change of $-C \cdot N$, i.e., it is equal and opposite to the change in the eigenvalue term, such that the total energy is not affected.

The equations above can be generalized to the spin-polarized case, as well. Because in the spin-polarized version of DFT there exist two potentials, $v_{\KS,\s}$, where $\s = \up$ or $\dn$, there also exist two sets of orbitals, ${\{\pphi_{i,\s}^{(\alpha)}\}}$, two densities, $n_\s$, and also two statistical weights $\alpha_\s$. As a result, Eq.~(\ref{eq.v_T}) is generalized to be
\begin{equation}
    v_{T,\s}(\rr) = \lp(\frac{\delta E_T}{\delta n_\s} \rp)_{n_\tau} = \lp( \frac{\de E_T}{\de \alpha_\s} \rp)_{\substack{n_\s \\ n_\tau,\alpha_\tau}} + \lp( \frac{\delta E_T}{\delta n_\s(\rr)} \rp)_{\substack{\alpha_\s \\ n_\tau,\alpha_\tau}},
\end{equation}
where $\tau$ refers to the other spin channel than $\s$. Other equations of this section can be generalized accordingly; In particular, Eq.(\ref{eq.v_T.0}) reads:
\begin{equation}
    v_{T,\s}^{(0)} = \lp( \frac{\de E_T}{\de \alpha_\s} \rp)_{\substack{\{\pphi_{i,\s}\} \\ \{\pphi_{i,\tau}\},\alpha_\tau}} - \int |\pphi_{N_0+1,\s}^{(\alpha_\s)}(\rr)|^2 v_{T,\s}^{(1)}(\rr) d^3r.
\end{equation}

We stress, however, that in all formalism presented in the main text and here only the $\alpha$ of one spin channel is allowed to be fractional.

\section{Numerical details on calculation of the systems H$_2$ and C}

The calculations presented in this work were performed using the DARSEC~\footnote{DARSEC stands for Diatomic All-electron Real-Space Electronic structure Calculations} code~\cite{Makmal09JCTC}. This program allows to perform spin-polarized all-electron DFT calculations for single atoms and diatomic molecules using the real-space approach~\cite{Chelikowsky94a,Chelikowsky94b,Kronik06} with a prolate-spheroidal grid~\cite{Wei,Laaksonen84,Kobus96,Makmal09JCTC}.

In all calculations the total energy $E$ and the highest occupied orbital energy, $\eps_{ho}$, were obtained within the numerical error of 1 mRy. The bond length of H$_2$ was found by relaxation to be 1.45 Bohr for LSDA (and therefore, by definition, also for eLSDA), and 1.39 Bohr for the EXX functional. This result is in close correspondence with previous calculations (see Ref.~\cite{Makmal09JCTC} and references therein). The bond length was kept unchanged when varying the number of electrons in the system.

Calculations with the orbital-dependent functionals, eLSDA and EXX, were performed using the OEP procedure exactly with the S-iteration method~\cite{KumPer03a,KumPer03b} and the Krieger-Li-Iafrate (KLI) approximation~\cite{KLI}, which is less demanding computationally
than a full OEP calculation. The differences between the total energy results obtained in the two methods were within the numerical accuracy of 1 mRy for eLSDA and within 4 mRy for EXX. This finding is consistent with a previous observation~\cite{KumPer03b} that KLI deviations from exact OEP results for ground-state energies are generally small. In any case, the remaining deviation of EXX results from the straight line condition is surely not just due to the use of KLI. This is because for the H$_2$ system, with (at most) one electron per spin channel, the OEP result is identical to the KLI one as the OEP orbital-shifts vanish, and yet deviation from linearity remains.

Furthermore, we note that incorporating the additional potential term $v^{(0)}$ introduced in this work (Eq.~(\eqmvzero) in the main text) in the OEP calculation does not require any other changes to the standard OEP procedure, or approximations thereof. Because $v^{(0)}$ is spatially uniform, it does not affect the KS orbitals and merely shifts the KS eigenvalues.

In the calculations performed for the C atom, it was assumed that the neutral C has a spin $S_z = 1$, the ion C$^+$ has the spin of $S_z = \half$, and for the ion C$^{++}$, $S_z = 0$. Therefore, varying the number of electrons in the system was performed solely in the spin-up channel. In addition, the axial quantum number $L_z$ was restricted to be 0 for $q=-2...-1$, and increased linearly with $q$ for $q=-1...0$, to obtain $L_z=1$ for the neutral C atom. Calculations with other values of $L_z$ were checked as well: the total energy they produced differed from the reported values by less than 4 mRy. These restrictions for the C system assured that the calculation is performed with an ensemble of two states, which was the one considered in the main text.

\bibliography{bibliography}